# Linear Fringe Field Effects of Quadrupoles

D. Zhou[1]

KEK National Laboratory, Tsukuba, Ibaraki 305-0801, Japan

J.Y. Tang[2], Y. Chen, N. Wang

Institute of High Energy Physics, CAS, Beijing 100049, China

**Abstract**

Fringe field becomes important when one requires more accurate modeling of a ring lattice to study the long-term beam dynamics in storage rings and deal with large aperture magnets in high-intensity proton synchrotrons or accumulator rings. In this paper, a simple expression to calculate the tune shifts due to quadrupole fringe fields is derived by using Lie algebra technique. With higher-order terms included, this method is more accurate compared with the linear fringe field model used in SAD code. The method is also applied to a BEPCII lattice. Also based on the Lie algebra technique and an inverse series technique, an equivalent hard-edge model for quadrupoles is proposed, in which the model parameters are derived analytically. The model has the advantages of the direct calculation of the equivalent length and strength of a quadrupole and the easy adaptation when the strength changes. The validity of the model and its simplified version has been checked with a numerical method, and they show very good agreements.

**Key words**: quadrupole fringe field, Lie algebra, tune shift, equivalent hard-edge model

**PACS Numbers**: 29.27.-a; 41.75.-i; 41.85.Lc; 29.20.Dh; 29.27.Bd

## I. INTRODUCTION

The effects of fringe fields in magnets have been investigated since a long time. With the fully developed analytical techniques such as Lie algebra [1] and differential algebra [2], it is convenient to calculate the linear and nonlinear maps for s-dependent magnetic fields such as

Corresponding authors:

[1]dmzhou@post.kek.jp;

[2]tangjy@ihep.ac.cn

quadrupole fringe fields [3-4]. The design of new light sources and colliders require more accurate modeling of the rings to study the long-term beam dynamics. For some proton synchrotrons of high intensity, large magnet apertures also make their fringe fields very important. There have been many studies on the linear and nonlinear effects of fringe field and the field interference between magnets, and much concern has been shown on the fringe field effects on beam dynamics, especially in small rings of a large acceptance [5].

Fringe field effects are often neglected when people carry out optics design and numerical simulations for large storage rings, where the fringe field extensions are much smaller than the magnet lengths. For these studies, the conventional hard-edge model [6] is usually chosen. That is, the focusing function of a quadrupole is approximated with a stepwise function. This is usually a good approximation for most studies even in small rings. In case that we need to take into account the field change along the optical axis, a more accurate description of the focusing function is to divide the magnet into many segments with each described by the hard-edge approximation [7]. The focusing function can be obtained from magnetic field measurements or 3D field numerical calculations [8] or analytical calculations [9]. For the convenience of particle tracking, the focusing functions of quadrupoles are often approximated by some specified functions [9-12]. Some of these functions reveal the relationships between the focusing function and the magnet parameters, such as length, aperture, etc. Once the focusing functions are known, linear and nonlinear maps can be calculated analytically or numerically.

With respect to the linear effect of quadrupole fringe field, G.E. Lee-Whiting presented an analytic approach using iterating integral equations derived from first-order differential equations [13]. Using this method, the transfer matrix of a quadrupole is calculated by convergent series containing form-factors calculated from field gradient integrals. Numerical checks show that no more than three terms of the series are needed to achieve a high accuracy modeling for a quadrupole with usual parameters. The effective lengths and strengths of quadrupoles are also derived from the form-factors explicitly. Lee-Whiting's method is straightforward and can be applied to any forms of focusing functions.

In Ref. [13], the transfer matrix of the fringe field region is directly calculated. To calculate lens parameters, quadrupole types with and without flat field regions should be treated differently. The corresponding formulae of matrix elements and lens parameters are a little complicated and the convergence properties are not quite good. In Irwin and Wang's work [3], fringe field is treated as perturbation on the ideal hard-edged lens and the fringe effects are calculated based on the Lie Algebra technique. Unlike the iterating integration method, the Lie Algebra technique preserves the symplectic properties of transfer matrices. Here, the fringe field is assumed to be anti-symmetric around the edge of a normal hard-edge model (abbr. as H.E. model), and the convergence properties of the method are quite good.

In this paper, we use the Lie Algebra technique similar to the one used by Irwin and Wang to analyze the linear effects of quadrupole fringe fields. The general description of fringe field properties is given in Section II. In Section III, an analytical expression of the tune shift due to quadrupole fringe fields is derived, and the application to a BEPCII (Beijing Electron-Positron Collider, phase II) lattice is presented. In Section IV, the expressions for an equivalent H.E. model for quadrupoles and its simplified version are derived, which is consistent to the one by other authors using an empirical method [7]; to check the validity of the model, the comparisons between the model and the numerical method are also given.

The fringe field effects of dipoles, the field overlapping or interference effects between magnets, and the nonlinear fringe field effects are beyond the scope of this paper.

## II. FRINGE FIELDS OF QUADRUPOLES

Conventional H.E. models are generally used in particle beam optics. For a quadrupole, the quadrupole coefficient and the length are defined as [6]

$$K_0 = \frac{eG_0}{p}, \tag{1}$$

$$L_0 = \frac{1}{G_0}\int_{-\infty}^{\infty} G(s)ds, \tag{2}$$

where $e$ is the particle's charge, $p$ is the particle's momentum, $G(s) = \left.\frac{\partial B_y}{\partial x}\right|_{x=y=0}$ is the on-axis field gradient, and $G_0$ is the peak value. Compared with a conventional H.E. model, a trapezoidal fringe model is more accurate to describe a real field distribution (Fig.1). The trapezoidal model is defined as

$$G_{tr}(s) = \begin{cases} \dfrac{G_0}{F_1}(s+s_0)+\dfrac{1}{2}G_0, & -s_0-\dfrac{1}{2}F_1 < s \leq -s_0+\dfrac{1}{2}F_1 \\ G_0, & -s_0+\dfrac{1}{2}F_1 < s \leq s_0-\dfrac{1}{2}F_1 \\ -\dfrac{G_0}{F_1}(s-s_0)+\dfrac{1}{2}G_0, & s_0-\dfrac{1}{2}F_1 < s \leq s_0+\dfrac{1}{2}F_1 \end{cases}, \quad (3)$$

where $s_0 = L_0/2$, and $F_1$ is the length of fringe field extension. Here we use the definition of SAD [14]:

$$F_1 = \sqrt{24 \left| \int_0^\infty \dfrac{\widetilde{G}(s)}{G_0}(s-s_0)ds \right|}, \quad (4)$$

where

$$\widetilde{G}(s) = \begin{cases} G(s)-G_0 & 0 < s < s_0 \\ G(s) & s \geq s_0 \end{cases}, \quad (5)$$

There are many fringe field functions used to model the smooth-gradient profile in the literature, and one of them is Enge function of 6 parameters [9]. The parameters of Enge function can be determined by fitting them with the data of magnetic field measurements. The definition of Enge function is:

$$E(s) = \dfrac{1}{1+\exp\left[a_1+a_2\left(\dfrac{s}{D_q}\right)+a_3\left(\dfrac{s}{D_q}\right)^2+a_4\left(\dfrac{s}{D_q}\right)^3+a_5\left(\dfrac{s}{D_q}\right)^4+a_6\left(\dfrac{s}{D_q}\right)^5\right]}, \quad (6)$$

where $D_q$ is the aperture of the magnet. The field gradient distribution described by Enge function is

$$G_{en}(s) = \begin{cases} G_0 E(-s-s_0), & s \leq 0 \\ G_0 E(s-s_0), & s > 0 \end{cases}, \quad (7)$$

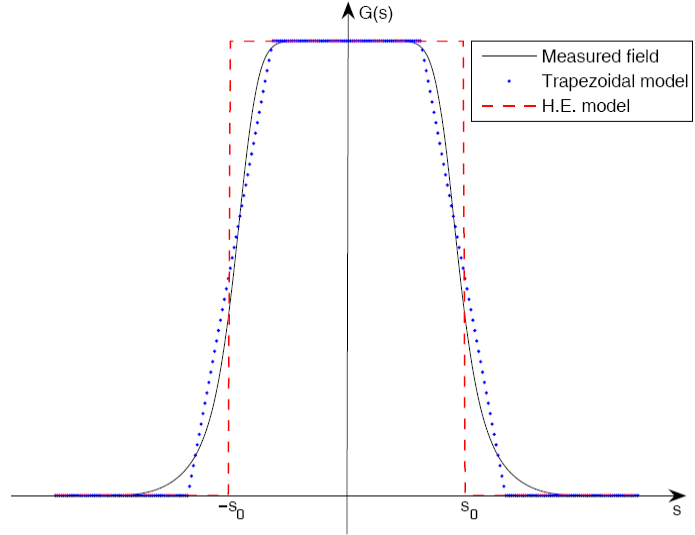

FIG.1. Conventional H.E. model and trapezoidal fringe model for a quadrupole

There is an implicit assumption that $E(-s_0) \to 0$ and $G_{en}(0) = G_0$. This is a good assumption for a magnet with long flat region in the body part. But when used to describe a full fringe field distribution, which has no flatted part, Eq. (7) is not appropriate [15]. In this case, functions such as bell-shape function or Gaussian function were proposed [13, 16]. The Gaussian function used here is

$$G_{ga}(s) = G_0 \exp(-\pi s^2 / d^2), \tag{8}$$

According to Eq. (2), the effective length of a field distribution described by Eq. (8) is $d$.

It should be mentioned that we have assumed that the shape of a fringe field does not change when the peak gradient changed. This means that the magnet is not in the status of saturation. Unfortunately, the shape of the fringe field usually changes due to magnet saturation.

### III. TUNE SHIFT DUE TO FRINGE FIELDS

#### A. Simple estimation of tune shift due to fringe field

The tune shift due to a quadrupole fringe field can be estimated with the theory of linear magnet imperfections [7]. For a distributed gradient error $k(s)$, the tune shift is

$$\Delta \nu = \frac{1}{4\pi} \oint \beta(s) k(s) ds, \tag{9}$$

where $\beta(s)$ is the beta function and $k(s)$ is the difference between the real field gradient and the one used in a traditional H.E. model. For the fringe field of a usual quadrupole, the quadrupole coefficient errors of the fringe field and its trapezoidal model with respect to its H.E. model are shown in Fig. 2. Assumed that the fringe field extension is small compared with the magnet length, the beta function near the magnet entrance/exit positions (defined by a traditional hard edge model, $s_0$ here) are approximated as

$$\beta(s) = \beta(s_0) - 2\alpha(s_0)(s - s_0) + \frac{1 + \alpha^2(s_0)}{\beta(s_0)}(s - s_0)^2, \quad (10)$$

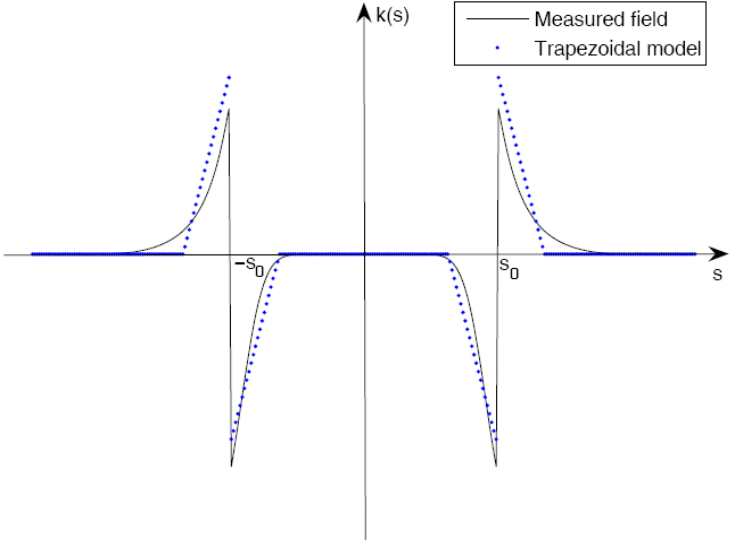

FIG. 2. Quadrupole coefficient errors of a real field distribution and its trapezoidal model with respect to the H.E. model

For convenience, we use the trapezoidal fringe model in the calculations instead of the real gradient distribution. The corresponding quadrupole coefficient error is

$$k_{tr}(s) = \begin{cases} 0, & 0 \le s \le s_0 - \frac{1}{2}F_1 \\ -\frac{K_0}{F_1}(s - s_0) - \frac{1}{2}K_0, & s_0 - \frac{1}{2}F_1 < s \le s_0 \\ -\frac{K_0}{F_1}(s - s_0) + \frac{1}{2}K_0, & s_0 < s \le s_0 + \frac{1}{2}F_1 \end{cases}, \quad (11)$$

Applying Eqs. (10) and (11) to Eq. (9), the tune shift due to the fringe fields of a quadrupole is expressed by the following approximation:

$$\Delta \nu = \frac{1}{48\pi} K_0 (\alpha_{in} - \alpha_{out}) F_1^2, \quad (12)$$

where $\alpha_{in}$ and $\alpha_{out}$ are the values of the $\alpha$ functions at the quadrupole entrance and exit, respectively. Eq. (12) holds in both the focusing and the defocusing planes. Here we notice that

$K_0$ is positive for the focusing plane and negative for the defocusing plane. Eq. (12) also reveals that the tune shift is determined by the magnet properties as well as the optics properties: 1) the quadrupole strength; 2) the fringe field extensions; 3) the beta function variation around the quadrupole entrance/exit.

### B. Linear fringe maps calculation based on Lie algebra technique

According to the theory of linear transformation, the coordinates of a charged particle crossing a quadrupole can be described by using differential equation

$$u'' \pm K(s)u = 0, \quad u = x, y, \tag{13}$$

where $K(s)$ is the on-axis quadrupole coefficient, $x$ and $y$ are the coordinates in the focusing and defocusing plane respectively. The relation of $K(s)$ and $G(s)$ is

$$K(s) = \frac{eG(s)}{p}, \tag{14}$$

The relevant Hamiltonian for Eq. (13) is

$$H(s) = \frac{1}{2}\left(p_x^2 + p_y^2\right) + \frac{1}{2}K(s)\left(x^2 - y^2\right), \tag{15}$$

Following the method developed by Irwin and Wang [3], the fringe fields of the quadrupole are treated as perturbations. Thus the Hamiltonian in the quadrupole can be divided into three parts: the entrance, the body and the exit. Then the s-dependent Hamiltonian (15) is rewritten as

$$H(s) = H_0(s) + \widetilde{H}(s), \tag{16}$$

where $H_0(s)$ and $\widetilde{H}(s)$ are the Hamiltonians of the body described by a H.E. model and the perturbation due to the fringe field. For the exit-side fringe field, we have:

$$H_0(s) = \begin{cases} \frac{1}{2}\left(p_x^2 + p_y^2\right) + \frac{1}{2}K_0\left(x^2 - y^2\right), & s_1 \leq s \leq s_0 \\ \frac{1}{2}\left(p_x^2 + p_y^2\right), & s_0 < s \leq s_2 \end{cases}, \tag{17}$$

and

$$\widetilde{H}(s) = \frac{1}{2}\widetilde{K}(s)\left(x^2 - y^2\right) = \begin{cases} \frac{1}{2}[K(s) - K_0]\left(x^2 - y^2\right), & s_1 \leq s \leq s_0 \\ \frac{1}{2}K(s)\left(x^2 - y^2\right), & s_0 < s \leq s_2 \end{cases}, \tag{18}$$

where $s_1 = 0$ is the center of the quadrupole, $s_2$ is a point far outside the fringe field region, and $s_0$ is the magnet end point defined by the H.E. model. The linear map from $s_1$ to $s_2$ can be written as

$$M(s_1 \to s_2) = R_-(s_1 \to s_0) R_+(s_0 \to s_2), \qquad (19)$$

$$R_-(s_1 \to s_0) = M_Q(s_1 \to s_0) e^{:f_2^-:}, \qquad (20)$$

$$R_+(s_0 \to s_2) = e^{:f_2^+:} M_{drift}(s_0 \to s_2), \qquad (21)$$

where

$$M_Q(s_1 \to s) \leftrightarrow \begin{bmatrix} \cos(\sqrt{K_0}s) & \frac{\sin(\sqrt{K_0}s)}{\sqrt{K_0}} & 0 & 0 \\ -\sqrt{K_0}\sin(\sqrt{K_0}s) & \cos(\sqrt{K_0}s) & 0 & 0 \\ 0 & 0 & \cosh(\sqrt{K_0}s) & \frac{\sinh(\sqrt{K_0}s)}{\sqrt{K_0}} \\ 0 & 0 & \sqrt{K_0}\sinh(\sqrt{K_0}s) & \cosh(\sqrt{K_0}s) \end{bmatrix}, \qquad (22)$$

$$M_{drift}(s_0 \to s) \leftrightarrow \begin{bmatrix} 1 & s-s_0 & 0 & 0 \\ 0 & 1 & 0 & 0 \\ 0 & 0 & 1 & s-s_0 \\ 0 & 0 & 0 & 1 \end{bmatrix}, \qquad (23)$$

Assumed that the perturbation is weak, the generating functions of the linear fringe maps can be calculated using the second-order BCH formulae:

$$f_2^- = -\int_{s_1}^{s_0} \overline{H}(s)ds + \frac{1}{2}\int_{s_1}^{s_0} ds \int_s^{s_0} ds' [\overline{H}(s), \overline{H}(s')], \qquad (24)$$

$$f_2^+ = -\int_{s_0}^{s_2} \overline{H}(s)ds + \frac{1}{2}\int_{s_0}^{s_2} ds \int_s^{s_2} ds' [\overline{H}(s), \overline{H}(s')], \qquad (25)$$

where

$$\overline{H}(s) = \begin{cases} \tilde{H}(s, M_Q(s_0 \to s)X), & s_1 \le s \le s_0 \\ \tilde{H}(s, M_{drift}(s_0 \to s)X), & s_0 < s \le s_2 \end{cases}, \qquad (26)$$

$$X = [x, p_x, y, p_y]^T, \qquad (27)$$

where $X$ is the phase space vector at $s_0$. To calculate $f_2^-$ and $f_2^+$ explicitly, Taylor expansion around $s_0$ to the third order is performed to the linear transfer matrix Eq. (22) and we obtain Eq. (28). Here we have assumed that the fringe field region is short.

$$M_Q(s_0 \to s) \leftrightarrow \begin{bmatrix} 1 - \frac{1}{2}K_0(s)\Delta s^2 & \Delta s - \frac{1}{6}K_0(s)\Delta s^3 & 0 & 0 \\ -K_0\left[\Delta s - \frac{1}{6}K_0(s)\Delta s^3\right] & 1 - \frac{1}{2}K_0(s)\Delta s^2 & 0 & 0 \\ 0 & 0 & 1 + \frac{1}{2}K_0(s)\Delta s^2 & \Delta s + \frac{1}{6}K_0(s)\Delta s^3 \\ 0 & 0 & K_0\left[\Delta s - \frac{1}{6}K_0(s)\Delta s^3\right] & 1 + \frac{1}{2}K_0(s)\Delta s^2 \end{bmatrix}, \quad (28)$$

where $\Delta s = s - s_0$. Then we can calculate the generating functions as

$$\begin{aligned} f_2^- &\cong -\frac{1}{2}I_0^-(x^2 - y^2) - I_1^-(xp_x - yp_y) - \frac{1}{2}I_2^-(p_x^2 - p_y^2) \\ &+ \frac{1}{2}K_0 I_2^-(x^2 + y^2) + \frac{2}{3}K_0 I_3^-(xp_x - yp_y) + \frac{1}{2}\Lambda_2^-(x^2 + y^2) \end{aligned} \quad (29)$$

$$\begin{aligned} f_2^+ &= -\int_{s_0}^{s_2} \overline{H}(s)ds + \frac{1}{2}\int_{s_0}^{s_2} ds \int_s^{s_2} ds' \left[\overline{H}(s), \overline{H}(s')\right] \\ &= -\frac{1}{2}I_0^+(x^2 - y^2) - I_1^+(xp_x - yp_y) - \frac{1}{2}I_2^+(p_x^2 - p_y^2) + \frac{1}{2}\Lambda_2^+(x^2 + y^2) \end{aligned} \quad (30)$$

From Eqs. (19-21), we also obtain the total linear fringe map

$$R_f = e^{:f_2^-:} e^{:f_2^+:} = e^{:f_2:}, \quad (31)$$

where

$$\begin{aligned} f_2 &\cong f_2^- + f_2^+ + \frac{1}{2}\left[f_2^-, f_2^+\right] \\ &\approx -(I_1^- + I_1^+)(xp_x - yp_y) - \frac{I_2^- + I_2^+}{2}(p_x^2 - p_y^2) \\ &+ \frac{K_0 I_2^-}{2}(x^2 + y^2) + \frac{2K_0 I_3^-}{3}(xp_x + yp_y) + \frac{\Lambda_2^- + \Lambda_2^+}{2}(x^2 + y^2), \\ &- \frac{1}{2}I_0^+(I_1^- + I_1^+)(x^2 + y^2) - \frac{1}{2}I_0^+(I_2^- + I_2^+)(xp_x + yp_y) \\ &+ \ldots \end{aligned} \quad (32)$$

The fringe field integrals (FFIs) are defined as

$$\begin{aligned} I_0^- &= \int_{s_1}^{s_0} \widetilde{K}(s)ds, & I_1^- &= \int_{s_1}^{s_0} \widetilde{K}(s)(s - s_0)ds \\ I_2^- &= \int_{s_1}^{s_0} \widetilde{K}(s)(s - s_0)^2 ds, & I_3^- &= \int_{s_1}^{s_0} \widetilde{K}(s)(s - s_0)^3 ds \end{aligned} \quad (33)$$

$$\begin{aligned} I_0^+ &= \int_{s_0}^{s_2} \widetilde{K}(s)ds, & I_1^+ &= \int_{s_0}^{s_2} \widetilde{K}(s)(s - s_0)ds \\ I_2^+ &= \int_{s_0}^{s_2} \widetilde{K}(s)(s - s_0)^2 ds, & I_3^+ &= \int_{s_0}^{s_2} \widetilde{K}(s)(s - s_0)^3 ds \end{aligned} \quad (34)$$

$$\Lambda_2^- = \int_{s_1}^{s_0} ds \int_s^{s_0} ds' \widetilde{K}(s)\widetilde{K}(s')(s'-s)$$
$$\Lambda_2^+ = \int_{s_0}^{s_2} ds \int_s^{s_2} ds' \widetilde{K}(s)\widetilde{K}(s')(s'-s)$$
(35)

Eqs. (33) and (34) are the FFIs from the zero-order to the third-order. Here the anti-symmetric assumption of fringe field is abandoned, though it is usually a good one [3]. From Eq. (32), the perturbation matrix in the focusing plane due to the exit fringe field is approximated as

$$M_{r,x} = \begin{bmatrix} 1 & 0 \\ J_3 & 1 \end{bmatrix} \begin{bmatrix} 1 & J_2 \\ 0 & 1 \end{bmatrix} \begin{bmatrix} e^{J_1} & 0 \\ 0 & e^{-J_1} \end{bmatrix},$$
(36)

where we define

$$J_1 = (I_1^- + I_1^+) - \frac{2K_0 I_3^-}{3} + \frac{1}{2} I_0^+ (I_2^- + I_2^+),$$
(37)

$$J_2 = I_2^- + I_2^+,$$
(38)

$$J_3 = K_0 I_2^- + (\Lambda_2^- + \Lambda_2^+) - I_0^+ (I_1^- + I_1^+),$$
(39)

In Eq. (32), the terms $xp_x \pm yp_y$ are the effects of image magnification elements, which correspond to the right matrix of the right side in Eq. (36). This is the leading term of the linear effect of fringe fields. The terms $x^2 + y^2$ are the effects of focusing elements, which correspond to the left matrix of the right side in Eq. (36). The terms $p_x^2 - p_y^2$ are the effects of drift elements, which correspond to the middle matrix of the right side in Eq. (36). The matrix form of Eq. (36) is symplectic.

Eq. (36) is similar to the treatment in SAD code on quadrupole fringe field. However, the focusing effect by $J_3$ is neglected in SAD.

The effect of the entrance-side fringe field can be calculated using symmetric conditions. Then we can obtain the whole focal length of an ordinary quadrupole as

$$f^{-1} = -T_{21} \cong \sqrt{K_0} \sin(\sqrt{K_0} L_0) e^{-2J_1} - 2J_3 \cos(\sqrt{K_0} L_0),$$
(40)

This formula is very simple compared with Eq. (27) in Ref. [13]. The correction on the focal length due to the fringe field is mainly contributed by the image magnification term.

Once the correction linear matrix is obtained, the tune shift due to the fringe field can be

calculated by using one-turn matrix of a ring:

$$\begin{bmatrix} \cos(2\pi Q_0) + \alpha \sin(2\pi Q_0) & \beta \sin(2\pi Q_0) \\ -\frac{1+\alpha^2}{\beta} \sin(2\pi Q_0) & \cos(2\pi Q_0) - \alpha \sin(2\pi Q_0) \end{bmatrix}_{out} M_{r,x}$$
$$= \begin{bmatrix} \cos(2\pi Q) + \alpha \sin(2\pi Q) & \beta \sin(2\pi Q) \\ -\frac{1+\alpha^2}{\beta} \sin(2\pi Q) & \cos(2\pi Q) - \alpha \sin(2\pi Q) \end{bmatrix}_{out}, \quad (41)$$

where $Q_0$ is the non-perturbated tune and $Q$ is the tune with the fringe field perturbation. From Eq. (41), the tune shift due to the exit-side fringe field can be calculated:

$$\Delta Q_{out} = Q - Q_0 \cong -\frac{\alpha_{out}}{2\pi} J_1 + \frac{1}{4\pi} \beta_{out} J_3 - \frac{1+\alpha_{out}^2}{4\pi \beta_{out}} J_2, \quad (42)$$

Compared with $J_1$, the terms containing $J_2$ and $J_3$ in Eq. (42) are higher-order terms and can be neglected in most cases. Similar result as Eq. (42) can be easily obtained for the tune shift due to the entrance-side fringe field. By comparing the definitions of $F_1$ and $J_1$, and keeping the leading term $I_1^- + I_1^+$ in $J_1$, the tune shift can be approximated by

$$\Delta Q_{out} \cong -\frac{\alpha_{out}}{2\pi} J_1 \cong -\frac{\alpha_{out} K_0 F_1^2}{48\pi}, \quad (43)$$

This means that the main part of the tune shift is proportional to the first-order FFIs, weighted by the local $\alpha$ function. Thus, Eq. (42) is consistent with Eq. (12) in the first order, but it contains higher-order terms.

### C. Tune shift due to fringe field at BEPCII

To check the validity of Eqs. (12) and (42), we choose a BEPCII synchrotron mode lattice and use SAD to calculate the optics parameters. This lattice was used during the BEPCII commissioning in 2007, and the nominal working point is (7.28, 5.18) [17]. There are seven types of quadrupoles in the BEPCII storage ring for synchrotron mode. For the type of 105Q, there are two sub-types with lengths of 0.31m and 0.34m, respectively. These two sub-types of quadrupoles have the same iron chamfer design, and the calculations and the field measurements show that the shapes of their fringe fields are very similar. The main parameters of these quadrupoles are listed in Table I. The fringe field lengths of the quadrupoles are calculated according to the magnetic measurements data by using Eq. (4).

TABLE I. Parameters of quadrupoles used in BEPCII synchrotron mode operation

| Quadrupole type | 105Q | 110Q | 160Q | Q1A | Q1B | Q2/Q3 | QSR |
|---|---|---|---|---|---|---|---|
| Effective length (m) | 0.310/0.340 | 0.408 | 0.646 | 0.254 | 0.464 | 0.548 | 0.240 |
| Aperture (mm) | 52.5 | 55.0 | 80.0 | 58.0 | 67.0 | 52.0 | 52.5 |
| Fringe field length (m) | 0.154 | 0.167 | 0.238 | 0.115 | 0.172 | 0.133 | 0.109 |
| Number | 44 | 10 | 6 | 2 | 2 | 4 | 1 |

For each quadrupole, the working point has been calculated with and without the fringe fields, thus the tune shift due to the fringe fields has been obtained by using Eq. (42). All the tune shifts are also compared with the calculations by using SAD where Eq. (12) is used, and shown in Fig. 3. The results agree well except at some strong focusing quadrupoles. When turning on the fringe fields of all the quadrupoles, the tune shift is calculated as (-0.036, -0.040).

The tune shifts of the whole ring were also measured with beam after the closed orbit correction but without the beta function correction. For the same lattice, the measured tune shift is about (-0.065, -0.090). This means that the tune shifts due to the fringe fields calculated by using SAD or Eq. (42) contribute only about half of the measured values. This can be explained by that the fringe fields of the bending magnets and the field interference between the quadrupoles and the sextupoles have also important contributions to the tune shift, which is confirmed by the studies on the magnetic field interference between quadrupoles and sextupoles [18].

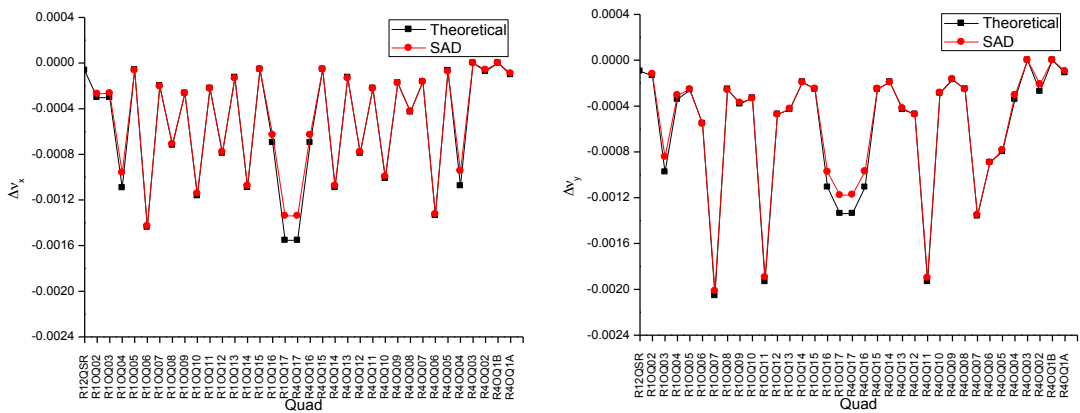

FIG. 3. Tune shifts due to the fringe fields of each quadrupole. Left: horizontal tune shift; Right: vertical tune shift. (Super period=2, half ring is shown)

## IV. EQUIVALENT HARD-EDGE MODEL

### A. Derivations by Lie Algebra Technique

By taking into account the linear effect of fringe fields, some people have proposed modified H.E. models [7, 16, 19]. The key point of these proposals is that the magnetic field is not the flat-top value of a real field distribution but the numerically calculated one based on the equivalent transfer matrix. With these modified H.E. models, they give different parameters for the focusing and defocusing planes. However, these numerical methods do not reveal the relation between the H.E. model parameters and the shape of fringe fields. In this section, an equivalent H.E. model is proposed, which is based on the Lie algebra technique and the perturbation method for the fringe fields.

Given that the focusing function $K(s)$ is symmetric around the center of a quadrupole magnet, the linear effect of the quadrupole including fringe fields can be always expressed by an equivalent H.E. model with a focusing strength $K_{eq}$ and a length $L_{eq}$. On the other hand, for any form of $K(s)$, the transfer matrix in the focusing plane of the whole magnet can be calculated numerically by using a slicing method (dividing the field distribution into slices) [6]. With the equivalent H.E. model, the transfer matrix is

$$M(-s_2 \to s_2) = \begin{bmatrix} T_{11} & T_{12} \\ T_{21} & T_{22} \end{bmatrix} = \begin{bmatrix} 1 & \lambda \\ 0 & 1 \end{bmatrix} \begin{bmatrix} \cos(\sqrt{K_{eq}}L_{eq}) & \frac{\sin(\sqrt{K_{eq}}L_{eq})}{\sqrt{K_{eq}}} \\ -\sqrt{K_{eq}}\sin(\sqrt{K_{eq}}L_{eq}) & \cos(\sqrt{K_{eq}}L_{eq}) \end{bmatrix} \begin{bmatrix} 1 & \lambda \\ 0 & 1 \end{bmatrix}, \quad (44)$$

where $\lambda = \dfrac{D_t - L_{eq}}{2}$, and $D_t$ is the total length chosen for the matrix calculation. Let $\alpha = \sqrt{K_{eq}}L_{eq}$, from Eq. (44) we obtain

$$\cos\alpha + \frac{1}{2}\alpha\sin\alpha = T_{11} - \frac{1}{2}D_t T_{21}, \quad (45)$$

$$L_{eq} = -\frac{\alpha\sin\alpha}{T_{21}}, \quad (46)$$

For a linear transfer system, there is always $T_{11}T_{22} - T_{12}T_{21} = 1$. If $K(s)$ is symmetric, the transfer matrix has an additional property of $T_{11} = T_{22}$. That is, there are only two free parameters in Eq. (44). Therefore, the solution of Eqs. (45-46) is unique accordingly.

Based on the perturbation method, the transfer matrix in Eq. (44) can be replaced by

$$M(-s_2 \to s_2) = \begin{bmatrix} 1 & \lambda \\ 0 & 1 \end{bmatrix} \begin{bmatrix} 1 & \Delta L \\ 0 & 1 \end{bmatrix} M_{r,x} \begin{bmatrix} \cos(\sqrt{K_0}L_0) & \frac{\sin(\sqrt{K_0}L_0)}{\sqrt{K_0}} \\ -\sqrt{K_0}\sin(\sqrt{K_0}L_0) & \cos(\sqrt{K_0}L_0) \end{bmatrix} M_{l,x} \begin{bmatrix} 1 & \Delta L \\ 0 & 1 \end{bmatrix} \begin{bmatrix} 1 & \lambda \\ 0 & 1 \end{bmatrix}, \quad (47)$$

where $M_{l,x}$ and $M_{r,x}$ are the perturbative matrices expressing the entrance and exit fringe field, respectively. And $\Delta L$ is half of the length difference between the traditional and equivalent H.E. models. $M_{l,x}$ has the following form:

$$M_{l,x} = \begin{bmatrix} e^{-J_1} & 0 \\ 0 & e^{J_1} \end{bmatrix} \begin{bmatrix} 1 & J_2 \\ 0 & 1 \end{bmatrix} \begin{bmatrix} 1 & 0 \\ J_3 & 1 \end{bmatrix}, \quad (48)$$

$$\Delta L = \frac{L_{eq} - L_0}{2}, \quad (49)$$

Equating Eqs. (47) and (44), we obtain

$$\begin{bmatrix} \cos(\sqrt{K_{eq}}L_{eq}) & \frac{\sin(\sqrt{K_{eq}}L_{eq})}{\sqrt{K_{eq}}} \\ -\sqrt{K_{eq}}\sin(\sqrt{K_{eq}}L_{eq}) & \cos(\sqrt{K_{eq}}L_{eq}) \end{bmatrix} = \begin{bmatrix} 1 & \Delta L \\ 0 & 1 \end{bmatrix} M_{r,x} \begin{bmatrix} \cos(\sqrt{K_0}L_0) & \frac{\sin(\sqrt{K_0}L_0)}{\sqrt{K_0}} \\ -\sqrt{K_0}\sin(\sqrt{K_0}L_0) & \cos(\sqrt{K_0}L_0) \end{bmatrix} M_{l,x} \begin{bmatrix} 1 & \Delta L \\ 0 & 1 \end{bmatrix}, \quad (50)$$

From Eq. (50), we can derive the following equations:

$$-\sqrt{K_0}\sin(\sqrt{K_0}L_0)e^{-2J_1} + 2J_3\cos(\sqrt{K_0}L_0) = -\sqrt{K_{eq}}\sin(\sqrt{K_{eq}}L_{eq}), \quad (51)$$

$$\cos(\sqrt{K_0}L_0) - \sqrt{K_0}\sin(\sqrt{K_0}L_0)e^{-2J_1}(\Delta L + J_2)$$
$$+ \frac{J_3 e^{2J_1}\sin(\sqrt{K_0}L_0)}{\sqrt{K_0}} + 2J_3\cos(\sqrt{K_0}L_0)\Delta L = \cos(\sqrt{K_{eq}}L_{eq}) \quad (52)$$

One can certainly solve Eqs. (51-52) numerically to obtain $K_{eq}$ and $L_{eq}$, however, it is also interesting to derive a solution expressed in expanded series. This can be obtained by using an inverse series method. More details are given in the appendix.

According to Eqs. (37-39), let

$$2J_1 = A \cdot K_0 + D \cdot K_0^2 \quad J_2 = B \cdot K_0 \quad J_3 = C \cdot K_0^2. \quad (53)$$

The parameters $A$, $B$, $C$ and $D$ are determined only by the shape of the fringe field, and they are not influenced by $K_0$ and $L_0$. Comparing with the definitions of $J_1$ and $F_1$, we find that $A$ is equal to $F_1^2/12$ when second and higher order FFIs can be neglected.

Following the inverse series method and the result for $\alpha$ in the appendix, the solutions for $K_{eq}$, $L_{eq}$ and $K_{eq}L_{eq}$ after a reasonable truncation to the series can be obtained:

$$\begin{aligned}K_{eq} \cong K_0 &\left[1+\left(-\frac{6A}{L_0^2}-\frac{12B}{L_0^3}+\frac{54A^2}{L_0^4}+\frac{216AB}{L_0^5}+\frac{216B^2}{L_0^6}-\frac{540A^3}{L_0^6}-\frac{3240A^2B}{L_0^7}+\frac{5670A^4}{L_0^8}\right)\right.\\ &+\left(\frac{2A}{5L_0^2}+\frac{24B}{5L_0^3}-\frac{12A^2}{L_0^4}-\frac{6D}{L_0^4}-\frac{72AB}{L_0^5}-\frac{36AC}{L_0^5}-\frac{108B^2}{L_0^6}-\frac{96BC}{L_0^6}\right.\\ &\left.+\frac{108AD}{L_0^6}+\frac{216BD}{L_0^7}+\frac{162A^3}{L_0^6}+\frac{1188A^2B}{L_0^7}+\frac{648A^2C}{L_0^7}-\frac{2079A^4}{L_0^8}\right)K_0L_0^2\\ &+\left(\frac{2A}{175L_0^2}+\frac{4B}{175L_0^3}+\frac{74A^2}{175L_0^4}+\frac{2D}{5L_0^4}+\frac{576AB}{175L_0^5}+\frac{24AC}{5L_0^5}+\frac{3096B^2}{175L_0^6}+\frac{136BC}{5L_0^6}+\frac{12C^2}{L_0^6}\right.\\ &\left.\left.-\frac{24AD}{L_0^6}-\frac{72BD}{L_0^7}-\frac{36CD}{L_0^7}-\frac{2434A^3}{175L_0^6}-\frac{23004A^2B}{175L_0^7}-\frac{162A^2C}{L_0^7}+\frac{18999A^4}{70L_0^8}\right)K_0^2L_0^4\right]\end{aligned}$$, (54)

$$\begin{aligned}L_{eq} \cong L_0 &\left[1+\left(\frac{6A}{L_0^2}+\frac{12B}{L_0^3}-\frac{18A^2}{L_0^4}-\frac{72AB}{L_0^5}-\frac{72B^2}{L_0^6}+\frac{108A^3}{L_0^6}+\frac{648A^2B}{L_0^7}-\frac{810A^4}{L_0^8}\right)\right.\\ &+\left(-\frac{2A}{5L_0^2}-\frac{14B}{5L_0^3}-\frac{2C}{L_0^3}+\frac{21A^2}{5L_0^4}+\frac{6D}{L_0^4}+\frac{24AB}{5L_0^5}+\frac{24AC}{L_0^5}+\frac{24B^2}{5L_0^6}+\frac{72BC}{L_0^6}\right.\\ &\left.-\frac{36AD}{L_0^6}-\frac{72BD}{L_0^7}-\frac{18A^3}{L_0^6}-\frac{72A^2B}{L_0^7}-\frac{180A^2C}{L_0^7}+\frac{162A^4}{L_0^8}\right)K_0L_0^2\\ &+\left(-\frac{2A}{175L_0^2}-\frac{4B}{175L_0^3}-\frac{A^2}{5L_0^4}-\frac{2D}{5L_0^4}-\frac{4AC}{L_0^5}-\frac{68BC}{5L_0^6}-\frac{12C^2}{L_0^6}\right.\\ &\left.\left.+\frac{42AD}{5L_0^6}+\frac{24BD}{5L_0^7}+\frac{24CD}{L_0^7}+\frac{53A^3}{35L_0^6}+\frac{24A^2B}{35L_0^7}+\frac{216A^2C}{5L_0^7}-\frac{663A^4}{70L_0^8}\right)K_0^2L_0^4\right]\end{aligned}$$, (55)

$$\begin{aligned}K_{eq}L_{eq} \cong K_0L_0 &\left[1+\left(\frac{2B}{L_0^3}-\frac{2C}{L_0^3}-\frac{3A^2}{L_0^4}-\frac{12AB}{L_0^5}-\frac{12B^2}{L_0^6}+\frac{108A^2B}{L_0^7}+\frac{18A^3}{L_0^6}-\frac{135A^4}{L_0^8}\right)K_0L_0^2\right.\\ &\left.+\left(\frac{A^2}{5L_0^4}+\frac{4AB}{5L_0^5}+\frac{24B^2}{5L_0^6}+\frac{4BC}{L_0^6}-\frac{6AD}{L_0^6}-\frac{12BD}{L_0^7}-\frac{3A^3}{L_0^6}-\frac{24A^2B}{L_0^7}-\frac{18A^2C}{L_0^7}+\frac{36A^4}{L_0^8}\right)K_0^2L_0^4\right]\end{aligned}$$, (56)

$K_{eq}$ and $L_{eq}$ are expressed by the values ($K_0$ and $L_0$) used in the traditional H.E. model and correction factors that are dependent on the fringe field integrals and $K_0$ and $L_0$ values. Although $K_0L_0^2$ can be larger than unit, the expression using $K_0L_0^2$ and $K_0^2L_0^4$ is still reasonable due to the intrinsic nature of the $K_{eq}$, $L_{eq}$ and $K_{eq}L_{eq}$ when the fringe field can be considered as a perturbation to the body field. In the square brackets of Eqs. (54) and (55), there is a constant term for $K_{eq}$ and $L_{eq}$, which is independent from $K_0$; however, there is no such constant term for $K_{eq}L_{eq}$.

For the defocusing plane, one needs only to change $K_0$ to $-K_0$. Therefore, the correction terms in Eqs. (54-56) are different from the ones in the focusing plane, and this is consistent with the results in Refs. [7, 16, 19].

## B. Comparisons between the equivalent hard-edge model and numerical calculations

In this section, we check the validity of the equivalent H.E. model by comparing its results with numerical calculations. For the convenience of calculation and comparison, we use a field contribution described by an Enge function as shown in Eq. (6) with the default parameters used in COSY INFINITY for quadrupoles [20-21]:

$a_1$=0.296471, $a_2$= 4.533219, $a_3$=-2.270982, $a_4$=1.068627, $a_5$=-0.036391, $a_6$=0.022261.

Firstly, we calculate the transfer matrices numerically using a slicing method to a quadrupole. Then we can obtain the equivalent H.E. model parameters by directly solving Eq. (43). Lastly, for different variations of the quadrupole parameters, the theoretical results based on Eqs. (54-56) are compared with the numerical results. The comparisons are also made for a simplified H.E. model as shown in Eqs. (57-59), which takes use only the constant and $K_0$ terms for the corrections.

$$K_{eq} \approx K_0 \left[ 1 + \left( -\frac{6A}{L_0^2} + \frac{54A^2}{L_0^4} - \frac{12B}{L_0^3} \right) + \frac{2A}{5} K_0 \right], \quad (57)$$

$$L_{eq} \approx L_0 \left[ 1 + \left( \frac{6A}{L_0^2} - \frac{18A^2}{L_0^4} + \frac{12B}{L_0^3} \right) - \frac{2A}{5} K_0 \right], \quad (58)$$

$$K_{eq} L_{eq} \approx K_0 L_0 \left[ 1 + \left( -\frac{3A^2}{L_0^2} + \frac{2B}{L_0} - \frac{2C}{L_0} \right) K_0 \right], \quad (59)$$

This simplified equivalent H.E. model is consistent to the empirical formulae in Ref. [7], which is accurate enough in most cases.

For a quadrupole with usual parameters, the equivalent H.E. model and its simplified version agree well. This can be seen in Table II where some of the BEPCII quadrupoles are used for the checking. The accuracies on $L_{eq}$ and $K_{eq}$ are in the order of 0.5 mm and 0.002 m$^{-2}$, respectively.

TABLE II. Applying the equivalent H.E. model to the BEPCII quadrupoles

| Quadrupole type | 105Q | 110Q | 160Q | Q1A | Q1B |
|---|---|---|---|---|---|
| Aperture (mm) | 52.5 | 55.0 | 80.0 | 58.0 | 67.0 |
| $L_0$ (m) | 0.310 | 0.408 | 0.646 | 0.254 | 0.464 |
| $K_0$ (max.,m$^{-2}$) | 1.51 | 1.33 | 0.54 | 0.83 | 0.67 |

| $A$ ($\times 10^{-3}$) | 1.893 | 2.487 | 5.124 | 1.159 | 2.126 |
|---|---|---|---|---|---|
| $B$ ($\times 10^{-5}$) | 4.15 | 5.18 | 11.5 | -0.199 | 3.16 |
| $C$ ($\times 10^{-5}$) | -1.65 | -2.87 | -9.87 | -1.19 | -2.46 |
| $D$ ($\times 10^{-7}$) | -0.354 | -0.635 | -43.5 | -2.53 | -3.86 |
| $L_{eq}$ (F/D, m$^{-2}$) [1] | 0.3489/0.3497 | 0.4459/0.4471 | 0.6943/0.6958 | 0.2796/0.2798 | 0.4921/0.4927 |
| $K_{eq}$ (F/D, m$^{-2}$) [1] | 1.3420/1.3378 | 1.2174/1.2133 | 0.5026/0.5013 | 0.7540/0.7536 | 0.6318/0.6310 |
| $L_{eq}$ (F/D, m$^{-2}$) [2] | 0.3493/0.3500 | 0.4461/0.4472 | 0.6944/0.6959 | 0.2794/0.2796 | 0.4922/0.4927 |
| $K_{eq}$ (F/D, m$^{-2}$) [2] | 1.3397/1.3362 | 1.2164/1.2129 | 0.5024/0.5013 | 0.7565/0.7559 | 0.6317/0.6309 |

Note: 1 denotes the equivalent H.E. model; 2 denotes the simplified equivalent H.E. model.

Then four cases are chosen to study the relationship between the equivalent H.E. model parameters and the parameters of a field distribution:

1) Case 1: the effective strength $K_0$ is taken as a variable and the shape of fringe field is fixed. In this case, the effective length $L_0$ and the aperture $D_q$ are chosen as 0.340 m and 0.105 m, which are from one of the BEPCII quadrupoles. The relative changes in quadrupole coefficient, equivalent length and focusing strength with different models are compared in Fig. 3. The agreement between the equivalent H.E. model and the numerical method is very good, and the accuracy by the simplified H.E. model is relatively worse but still a good approximation for usual $K_0 L_0$ values ($K_0 L_0 < 2$). As indicated in Eqs. (54-55), the constant deviations on $\Delta K / K_0$ and $\Delta L / L_0$ from the conventional H.E. model ($K_0 L_0$) for both the focusing and defocusing planes are identical. Since the parameters of the equivalent H.E. model are derived from the transfer matrix elements $T_{11}$ and $T_{21}$, the relative changes on $T_{11}$ and $T_{21}$ between the numerical method and the equivalent H.E. model are compared with those calculated from Eq. (47), and are shown in Fig. 4. The agreements in the transfer matrix elements between the equivalent H.E. model and the numerical method are quite good within a very large range of $K_0$, in the order of $5 \times 10^{-4}$. They are comparable to the results by directly using Eq. (47) which gives an agreement in the order of $2 \times 10^{-4}$.

2) Case 2: the magnet aperture $D_q$ is taken as a variable. In this case, the effective strength $K_0$ and the effective length $L_0$ are fixed as 2.0 m$^{-2}$ and 0.6 m, respectively. The relative changes in quadrupole coefficient, equivalent length and focusing strength with different models are compared in Fig. 5, and the relative changes on the transfer matrix elements between the numerical method and the equivalent H.E. model are shown in Fig. 6. The fringe field length $f_1$ calculated by using Eq. (4) instead of $D_q$ is chosen as the abscissa in the figures. We can see quite important changes in the Equivalent H.E. model parameters with relatively large $f_1$ values, and this is reasonable because a larger aperture changes the field distribution largely. The agreement between the equivalent H.E. model and the numerical method is even better compared with the variation of $K_0$ when $f_1$ is smaller than $L_0/2$, and the same conclusion is also valid for the simplified H.E. model. However, the accuracy of the equivalent H.E. model becomes worse when $f_1$ becomes close to $L_0$. Comparing with the results calculated from Eq. (47), we can conclude that a large fringe field extension degrades the convergence properties of the series solutions of Eqs. (51-52). This will be studied further in Case 4 where a full fringe field distribution is chosen.

3) Case 3: the effective length $L_0$ is taken as a variable. In this case, the profile of the fringe field does not change and the effective strength $K_0$ is set as 10.0 m$^{-2}$. The aperture $D_q$ is fixed as 0.105 m. A relatively large $K_0$ is chosen here for the purpose of checking the validity of the models in extreme cases. The relative changes in quadrupole coefficient, equivalent length and focusing strength with the different models are compared in Fig. 7, and the relative changes in the transfer matrix elements between the numerical method and the equivalent H.E. model are shown in Fig. 8. There is a reverse tendency comparing with the variation in $f_1$; in other words, with a larger $L_0$ or a smaller $f_1$ the equivalent H.E. model will approach the conventional H.E. model. The agreement in the transfer matrix elements between the results calculated from Eq. (47) and the numerical method is better than $1\times10^{-4}$ even in this case of a very large $K_0$.

4) Case 4: to check the effectiveness of the equivalent H.E. model applying to a full fringe field distribution, the field distribution is described by a Gaussian function as shown in Eq. (8). In this case, the effective strength $K_0$ is set as 2.0 m$^{-2}$. The relative changes on the transfer matrix elements between the numerical method and the equivalent H.E. model are shown in Fig. 9. From the comparison, we can see that the equivalent H.E. model is much less accurate when applying it to a quadrupole with full fringe field. This is mainly due to the truncation errors as discussed in the appendix. When fringe field length is comparable to $L_0$, the higher order terms of $A$ can not be neglected in Eqs. (54-56). This is always true for the case of full fringe field distribution, even when $L_0$ is very small. That is why there is a relative shift on $T_{21}$ in Fig.9b. The bad result at $L_0$ close to 0.4 is due that $T_{11}$ is close to zero. For the comparison, the solution from Eq. (47) still gives good accuracy, in the order of $10^{-4}$.

For all the four cases shown above, we find that a quadrupole field distribution with a large $K_0$, a smaller $L_0$ or a large fringe field extension will degrade the accuracy of the equivalent H.E.

model and to a less degree of the solution from Eq. (47). In case of a full fringe field distribution with a large $L_0$, higher accuracy requires more higher-order terms of FFIs in Eq. (47) to be included, and more higher-order terms in $K_0$ and more higher-order coefficients in $A$, $B$, $C$, $D$ are needed.

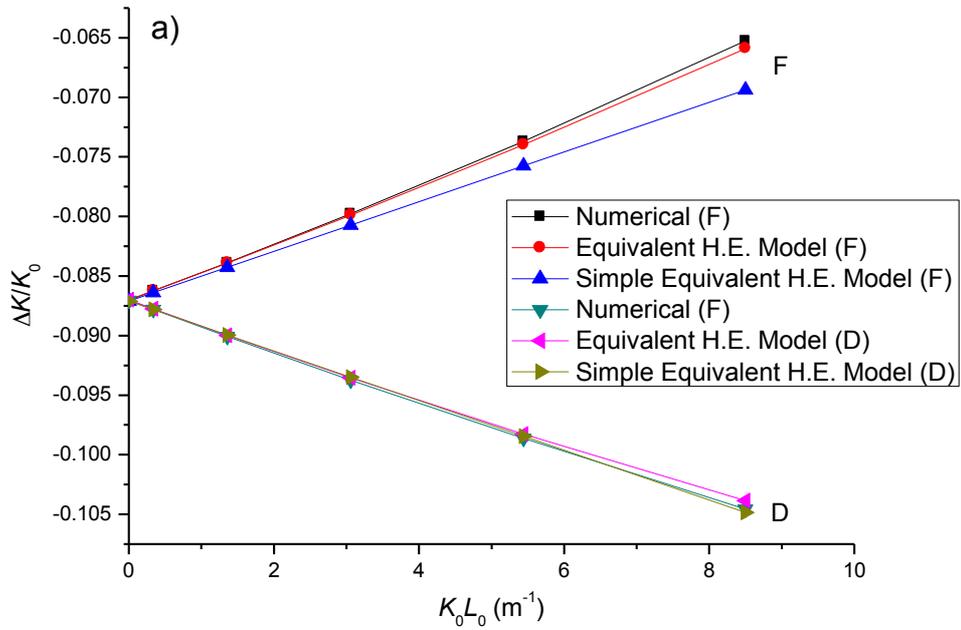

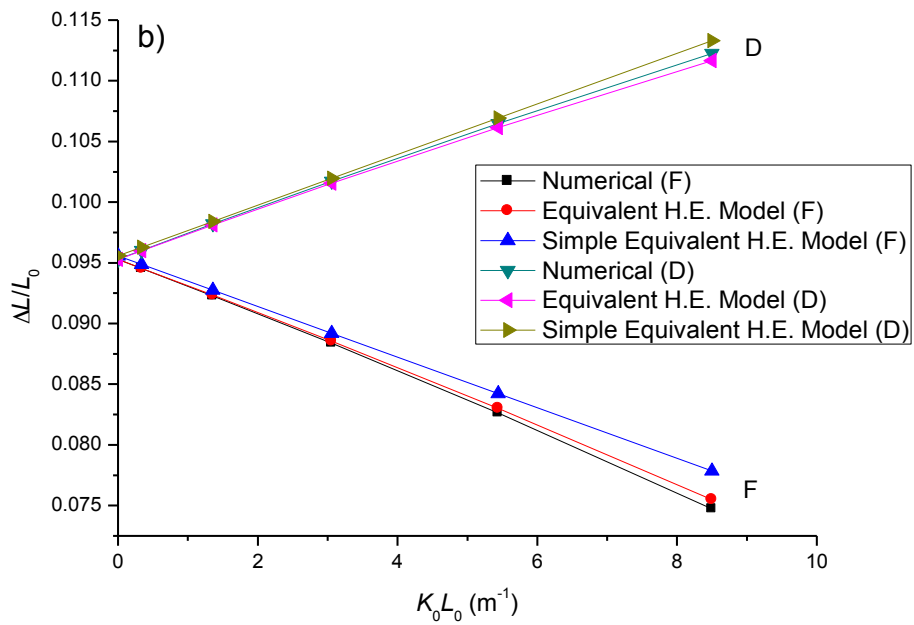

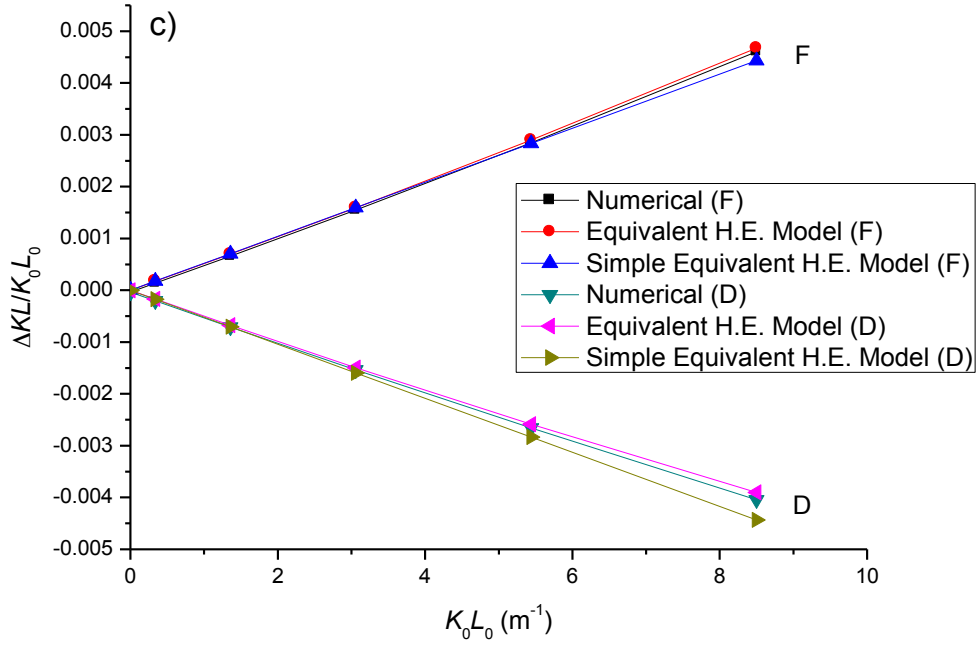

FIG. 3. Relative changes in quadrupole coefficient (a), equivalent length (b) and focusing strength (c) dependent on $K_0L_0$ among different models (only $K_0$ varies). F and D are for the focusing and defocusing planes, respectively.

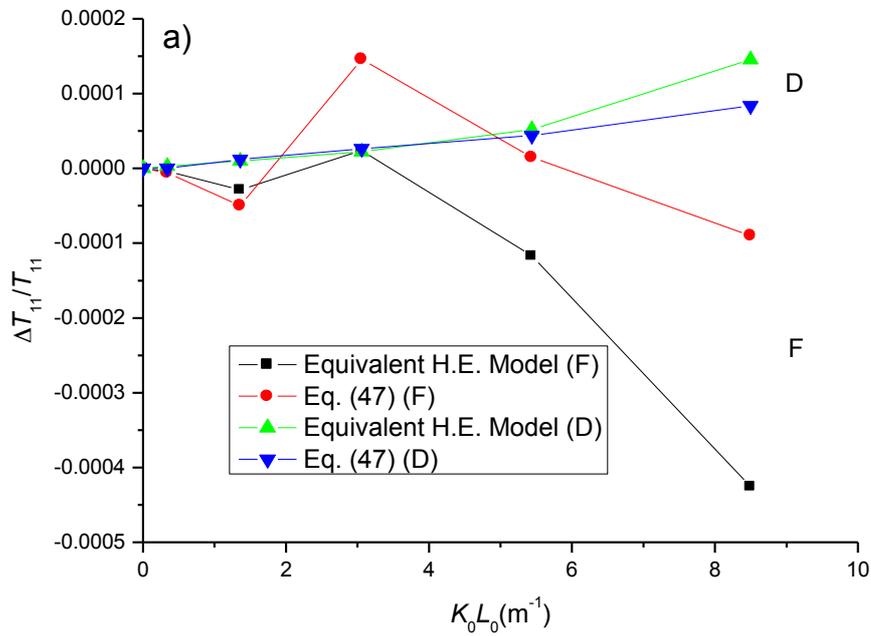

FIG. 4. Relative changes in the transfer matrix elements $T_{11}$ (a) and $T_{21}$ (b) dependent on $K_0L_0$ by comparing the equivalent H.E. model and the solution of Eq. (47) with the numerical method.

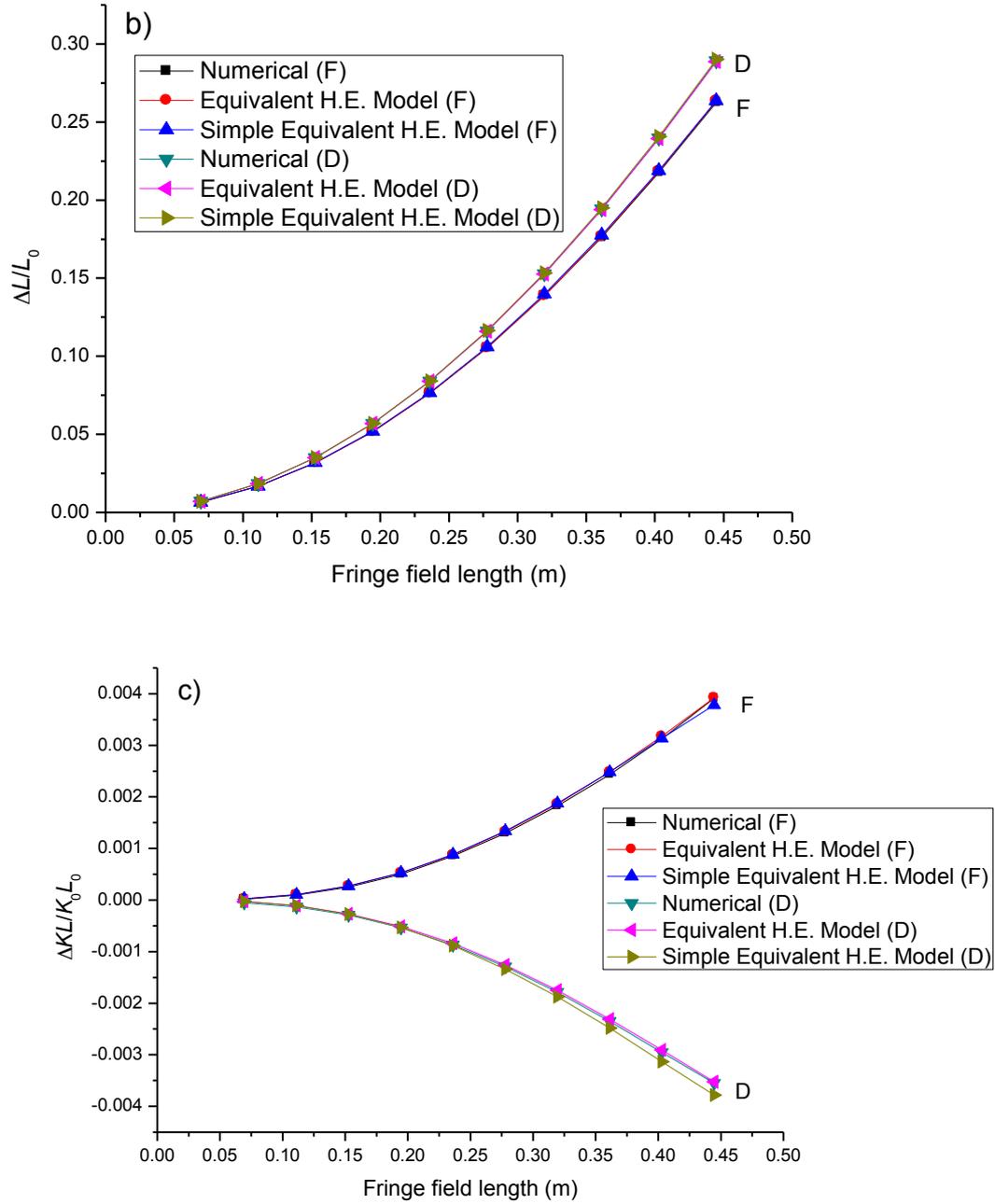

FIG. 5. Relative changes in quadrupole coefficient (a), equivalent length (b) and focusing strength (c) dependent on the fringe field length among the different models. F and D are for the focusing and defocusing planes, respectively.

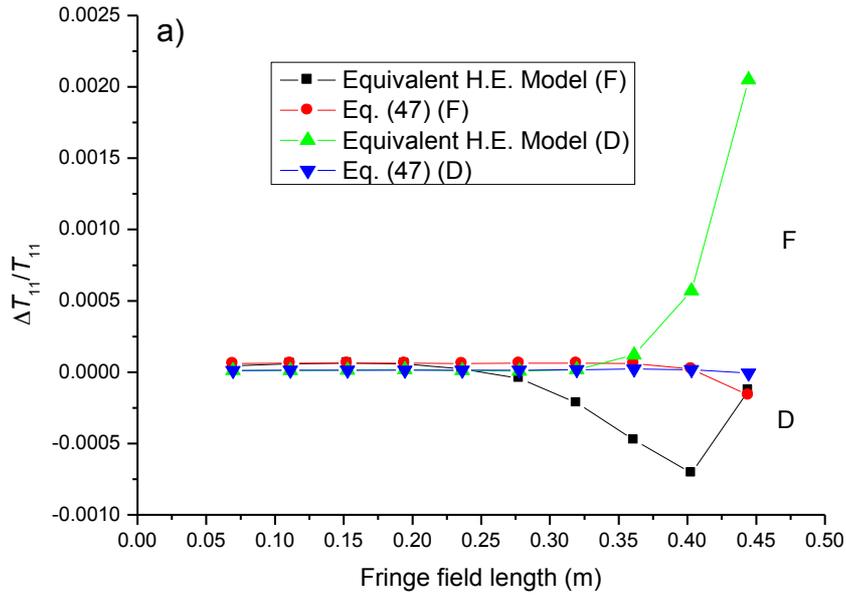

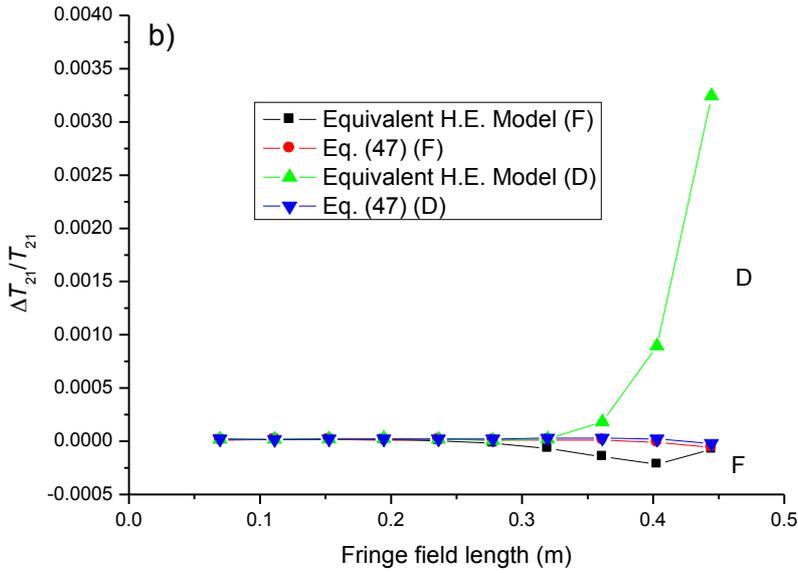

FIG. 6. Relative changes in the transfer matrix elements $T_{11}$ (a) and $T_{21}$ (b) dependent on the fringe field length by comparing the equivalent H.E. model and the solution of Eq. (47) with the numerical method.

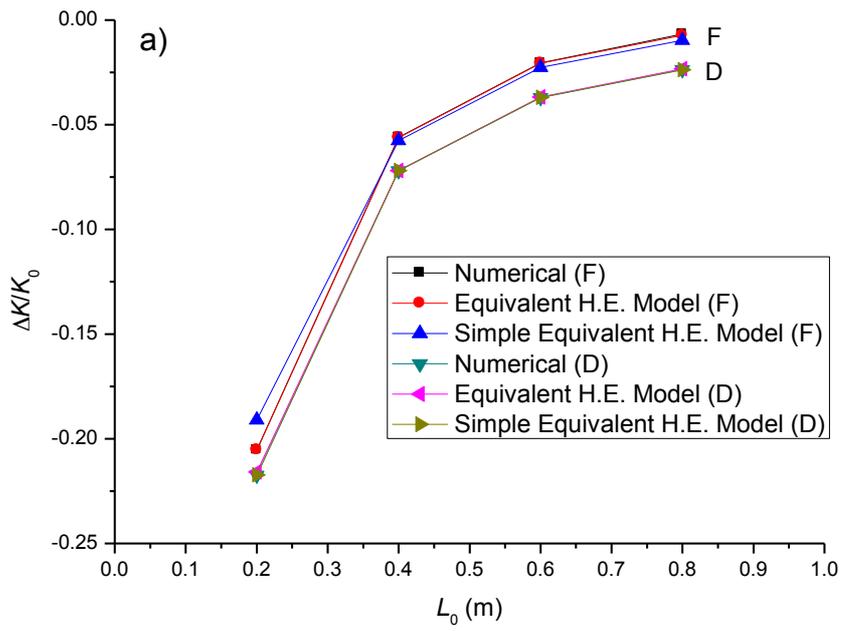
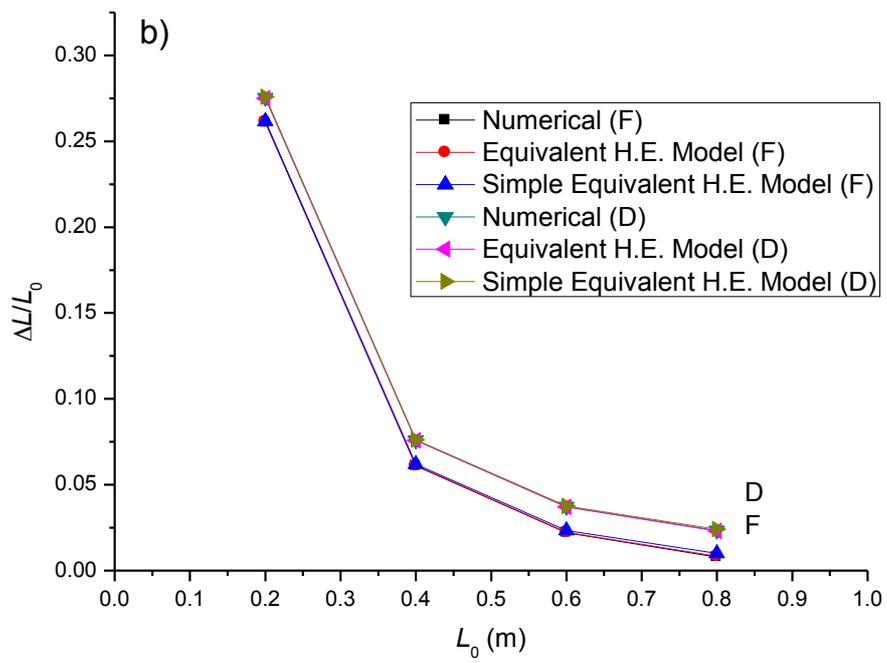

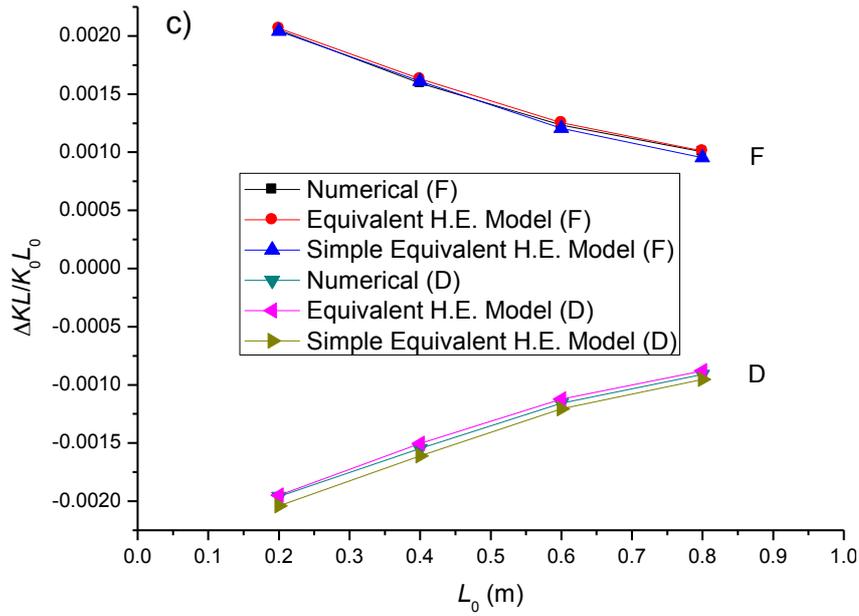

FIG. 7. Relative changes in quadrupole coefficient (a), equivalent length (b) and focusing strength (c) dependent on $L_0$ among different models. F and D are for the focusing and defocusing planes, respectively.

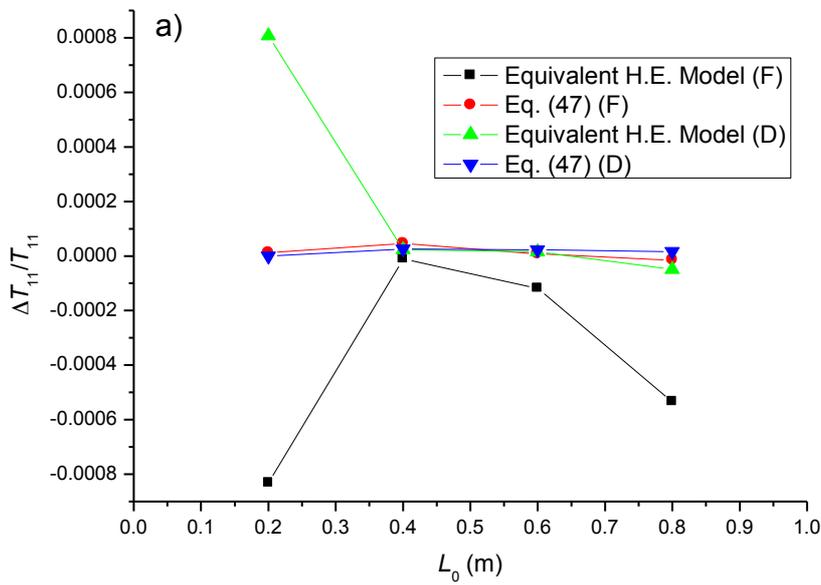

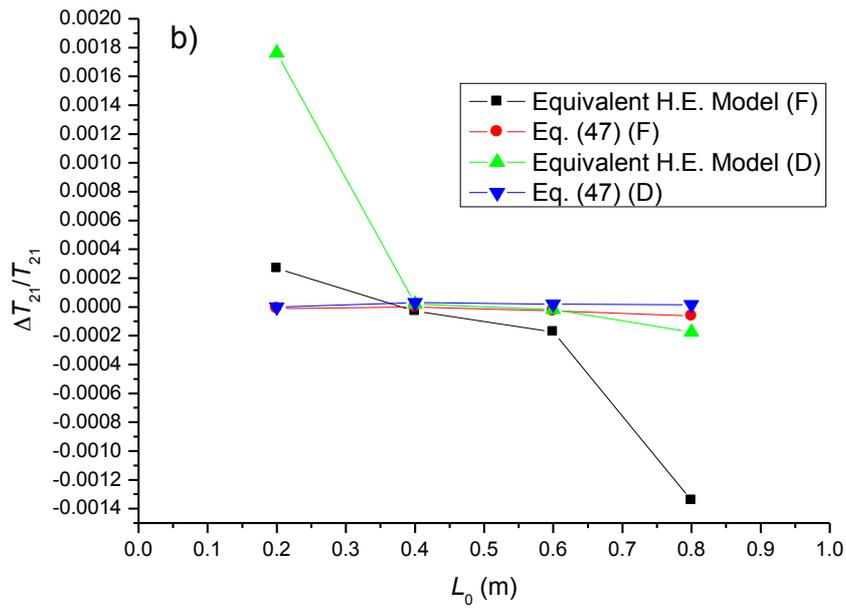

FIG. 8. Relative changes in the transfer matrix elements $T_{11}$ (a) and $T_{21}$ (b) dependent on $L_0$ by comparing the equivalent H.E. model and the solution of Eq. (47) with the numerical method

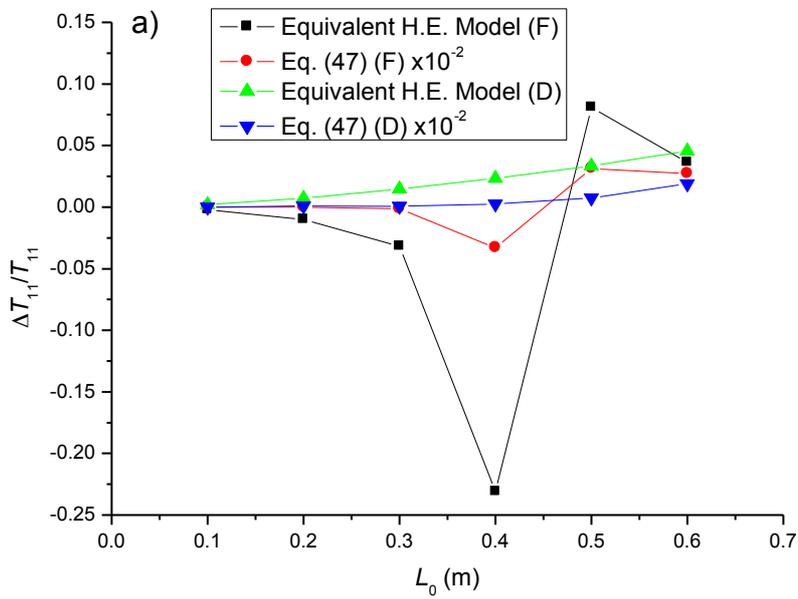

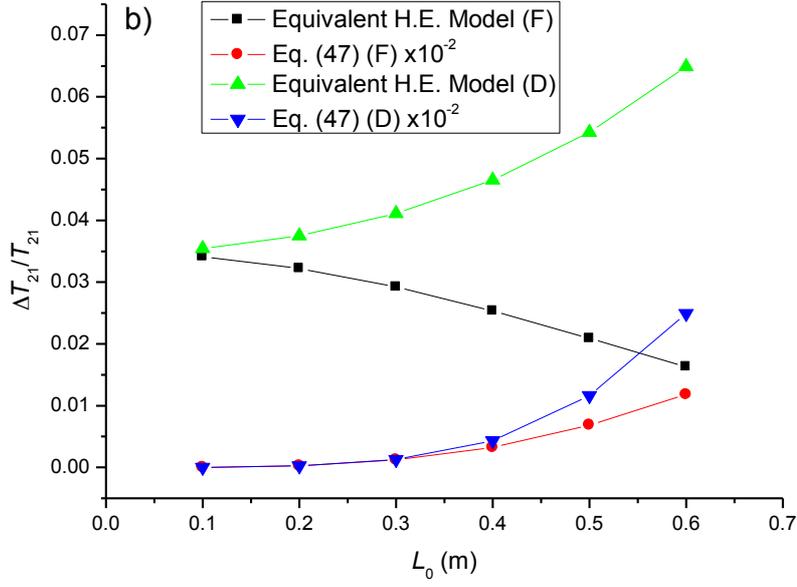

FIG. 9. Relative changes in the transfer matrix elements $T_{11}$ (a) and $T_{21}$ (b) dependent on the effective length of a full fringe-field quadrupole by comparing the equivalent H.E. model and the solution of Eq. (47) with the numerical method

### C. Discussion about the application of the equivalent H.E. model

In the last section, we have checked the validity of the equivalent H.E. model to a room-temperature quadrupole whose field distribution can be expressed by an Enge function. The model should be also applicable to a superconducting quadrupole whose fringe field distribution is more complicated, because only the FFIs play roles in the model.

What can we do with the equivalent H.E. model? Since most of the widely used linear optics codes don't take into the difference in the effective length and strength in the two transverse planes, it is not easy to implant the equivalent H.E. model. One possibility is to compute lattice functions separately in the two planes by using different magnet parameters. In this case, the conventional H.E. model is firstly used to determine $K_0$ and $L_0$ values, and then $K_{eq}$ and $L_{eq}$ can be applied. In the future, one can consider adding correction terms such as $\Delta K_F$, $\Delta K_D$, $\Delta L_F$ and $\Delta L_D$ to magnet elements in the codes.

## V. CONCLUSIONS

The tune shift due to the fringe fields of a quadrupole has been derived by using the Lie algebra technique, and it is compared with the linear fringe field model used in SAD by using the BEPCII lattice. They agree quite well in the first order, but the former consists of higher orders. Also based on the Lie algebra technique and the inverse series technique, an equivalent H.E. model that uses the derived model parameters. The validity of the model and its simplified version has been checked with a numerical method, and they show very good agreements for a quadrupole with usual parameters and less good agreement for a quadrupole with a very large aperture and a short magnet body.

The method described in this paper can be readily extended to describe the linear effect of a dipole and the nonlinear effect of a quadrupole or a dipole.

## ACKNOWLEDGEMENTS

This work was supported jointly by the National Natural Science Foundation of China (10775153) and the BEPCII project.

The authors would like to thank the BEPCII commissioning team for their great work in machine studies. Special thanks are due to C.T. Shi for providing the data of magnetic field measurements and the detailed discussions on the properties of the BEPCII quadrupoles. We would like to thank Q. Qin and G. Xu for the fruitful discussions.

# APPENDIX: INVERSE SERIES METHOD TO OBTAIN THE EQUIVALENT H.E. MODEL PARAMETERS

Combining Eq. (51) and Eq. (52), we have

$$\cos\alpha_0 + \frac{1}{2}\alpha_0 \sin\alpha_0 e^{-2J_1} - \frac{\alpha_0}{L_0}\sin\alpha_0 e^{-2J_1} J_2 + J_3 L_0 \left(\frac{e^{2J_1}\sin\alpha_0}{\alpha_0} - \cos\alpha_0\right)$$
$$= \cos\alpha + \frac{1}{2}\alpha\sin\alpha$$
(A-1)

where $\alpha_0 = \sqrt{K_0 L_0}$.

Let

$$F(\alpha) = \cos\alpha + \frac{1}{2}\alpha\sin\alpha,$$
(A-2)

Since $J_1$, $J_2$ and $J_3$ are small values, we can conclude that $\alpha$ equals to $\alpha_0$ plus perturbation terms of FFIs. From Eq. (A-2), we find a series for the inverse function of $F(\alpha)$ around $\alpha_0$:

$$\alpha = \alpha_0 + \frac{2\left(F - \cos\alpha_0 - \frac{1}{2}\alpha_0 \sin\alpha_0\right)}{\alpha_0 \cos\alpha_0 - \sin\alpha_0} + \frac{2\alpha_0 \sin\alpha_0 \left(F - \cos\alpha_0 - \frac{1}{2}\alpha_0 \sin\alpha_0\right)^2}{(\alpha_0 \cos\alpha_0 - \sin\alpha_0)^3}$$
$$+ O\left[\left(F - \cos\alpha_0 - \frac{1}{2}\alpha_0 \sin\alpha_0\right)^3\right]$$
(A-3)

According to the definitions in Eq. (53):

$$\begin{cases} A = 2(I_1^- + I_1^+)K_0 \\ B = (I_2^- + I_2^+)/K_0 \\ C = [K_0 I_2^- + (\Lambda_2^- + \Lambda_2^+) - I_0^+(I_1^- + I_1^+)]/K_0^2, \\ D = \left[-\frac{2K_0 I_3^-}{3} + \frac{1}{2}I_0^+(I_2^- + I_2^+)\right]/K_0^2 \end{cases}$$
(A-4)

Considering the property of a fringe field, there are:

$$\left|\widetilde{K}(s)\right| \le \frac{1}{2}K_0 \quad \text{and} \quad f_1 \le L_0$$

Assume $\varepsilon$ is a small number ($\varepsilon<1$), one can obtain the following relations according to Eqs. (33-35):

$$\begin{cases} -I_0^- \sim I_0^+ \sim \varepsilon K_0 L_0 \\ I_1^- \sim I_1^+ \sim \dfrac{1}{4}\varepsilon K_0 L_0^2 \\ -I_2^- \sim I_2^+ \sim \dfrac{1}{12}\varepsilon K_0 L_0^3, \\ I_3^- \sim I_3^+ \sim \dfrac{1}{32}\varepsilon K_0 L_0^4 \\ \Lambda_2^- \sim \Lambda_2^+ \sim \dfrac{1}{48}\varepsilon K_0^2 L_0^3 \end{cases} \tag{A-5}$$

Thus, one can obtain:

$$\frac{A}{L_0^2} \sim \varepsilon, \quad \frac{B}{L_0^3} \sim \frac{1}{12}\varepsilon, \quad -\frac{C}{L_0^3} \sim \frac{1}{6}\varepsilon, \quad -\frac{D}{L_0^4} \sim \frac{1}{96}\varepsilon, \tag{A-6}$$

In most cases, $\varepsilon < 0.1$; in the extremity case of a full fringe distribution, $\varepsilon \sim 0.25$. Therefore, in the expanded series, the terms can be truncated to $A^4, B^2, C^2, D^1$. One can see some practical examples of the coefficients $A$, $B$, $C$, $D$ in Table II.

Combining Eqs. (A-1) and (A-3), one can find the solution of $\alpha$ expressed by series of $K_0$, $L_0$, $A$, $B$, $C$ and $D$:

$$\begin{aligned}
\alpha = \alpha_0 \Bigg[ &1 + \left( \frac{6A}{L_0^2} + \frac{12B}{L_0^3} - \frac{18A^2}{L_0^4} - \frac{72AB}{L_0^5} + \frac{108A^3}{L_0^6} - \frac{72B^2}{L_0^6} + \frac{648A^2B}{L_0^7} - \frac{810A^4}{L_0^8} + \ldots \right) \\
&+ \Bigg( -\frac{2A}{5L_0^2} - \frac{14B}{5L_0^3} - \frac{2C}{L_0^3} + \frac{21A^2}{5L_0^4} + \frac{6D}{L_0^4} + \frac{24AB}{5L_0^5} + \frac{24AC}{L_0^5} \\
&\quad -\frac{18A^3}{L_0^6} + \frac{24B^2}{5L_0^6} + \frac{72BC}{L_0^6} - \frac{36AD}{L_0^6} - \frac{72A^2B}{L_0^7} - \frac{180A^2C}{L_0^7} - \frac{72BD}{L_0^7} + \ldots \Bigg) K_0 L_0^2 \\
&+ \Bigg( -\frac{2A}{175L_0^2} - \frac{4B}{175L_0^3} - \frac{2D}{5L_0^4} - \frac{A^2}{5L_0^4} + \frac{4AC}{L_0^5} + \frac{53A^3}{35L_0^6} - \frac{68BC}{5L_0^6} - \frac{12C^2}{L_0^6} + \frac{42AD}{5L_0^6} \\
&\quad + \frac{24A^2B}{35L_0^7} + \frac{216A^2C}{5L_0^7} + \frac{24BD}{5L_0^7} + \frac{24CD}{L_0^7} + \ldots \Bigg) K_0^2 L_0^4 + O\!\left[ (K_0 L_0^2)^3 \right] \Bigg]
\end{aligned} \tag{A-7}$$

Using the similar method, one can obtain the solutions of $K_{eq}$ and $L_{eq}$ from Eqs. (51-52).